\documentclass[conference]{IEEEtran}
\usepackage{cite}
\usepackage{amsmath,amssymb,amsfonts}
\usepackage{graphicx}
\usepackage{subfigure}
\usepackage{textcomp}
\usepackage{xcolor}
\usepackage{makecell}
\usepackage{stfloats}
\usepackage{gensymb}
\usepackage{multirow}
\usepackage{soul, color}
\usepackage{url}

\def\hrulefill{\leavevmode\leaders\hrule height 0.8pt\hfill\kern0pt}
\makeatletter
\def\rulefill{\leavevmode\leaders\hrule depth -3pt height 4pt\hfill\kern0pt}
\makeatother
\def\BibTeX{{\rm B\kern-.05em{\sc i\kern-.025em b}\kern-.08em
    T\kern-.1667em\lower.7ex\hbox{E}\kern-.125emX}}
\begin{document}

\title{PEMark:Watermarking API Responses Based on Proxy Gateways and Position Encoding}

\author
{\IEEEauthorblockN{Yifei Zhou$^{1}$, Xianjun Gu$^{1,2,*}$, Xinyu Dai$^{1}$, Ming Liu$^{1,3}$, Lansheng Han$^{*1,3}$}
\IEEEauthorblockA{
	\textit{$^{1}$Huazhong University of Science and Technology, Wuhan, China}\\
	\textit{$^{2}$State Grid Hubei Electric Power Co., Ltd., Wuhan, China}\\
	\textit{$^{3}$Wuhan Jinyinhu Laboratory, Wuhan, China}\\
	\text{LanshengHan@hust.edu.cn}
}}
\maketitle

\begin{abstract}
Data leakage from API responses has drawn wide attention. APIs are often not fully regulated, making them easy to abuse. One common solution is to embed watermarks into API responses for traceability. However, existing watermarking methods often require modifying database content or API response data. This forces changes to business system code, and may even disrupt normal business operations because data values are altered. In this paper, we propose an original pluggable watermarking scheme based on a watermark proxy gateway and PEMark (Position Encoding-based Watermarking). The key novelty of our approach is exploiting the inherent permutation redundancy in the ordering of JSON/XML key-value pairs---an overlooked dimension that carries no semantic information yet provides abundant encoding capacity. First, we forward server responses to the watermark proxy gateway, a design that requires zero modification to existing business systems. Then, we embed a watermark into each API response using position encoding, which reorders keys without altering any data values. To the best of our knowledge, this is the first work to achieve distortion-free API response watermarking via position encoding over a proxy gateway. Our method does not modify any data values, so normal business operations continue seamlessly after watermark embedding. Experimental results show that our framework maintains business usability while ensuring that returned API data is traceable. Compared with current mainstream schemes, our method is robust against tampering and insertion attacks (100\% similarity), and can withstand certain levels of deletion attacks.
\end{abstract}

\begin{IEEEkeywords}
API response watermarking, proxy gateway, position encoding, traceability, robustness.
\end{IEEEkeywords}

\section{Introduction}
The API (Application Programming Interface) is widely used for service provision, data acquisition, and AI model responses. However, when an API is not properly authorized or returns sensitive data, malicious calls may cause serious data leakage.

\begin{figure*}[!htbp]
	\centering
	\includegraphics[width=7 in]{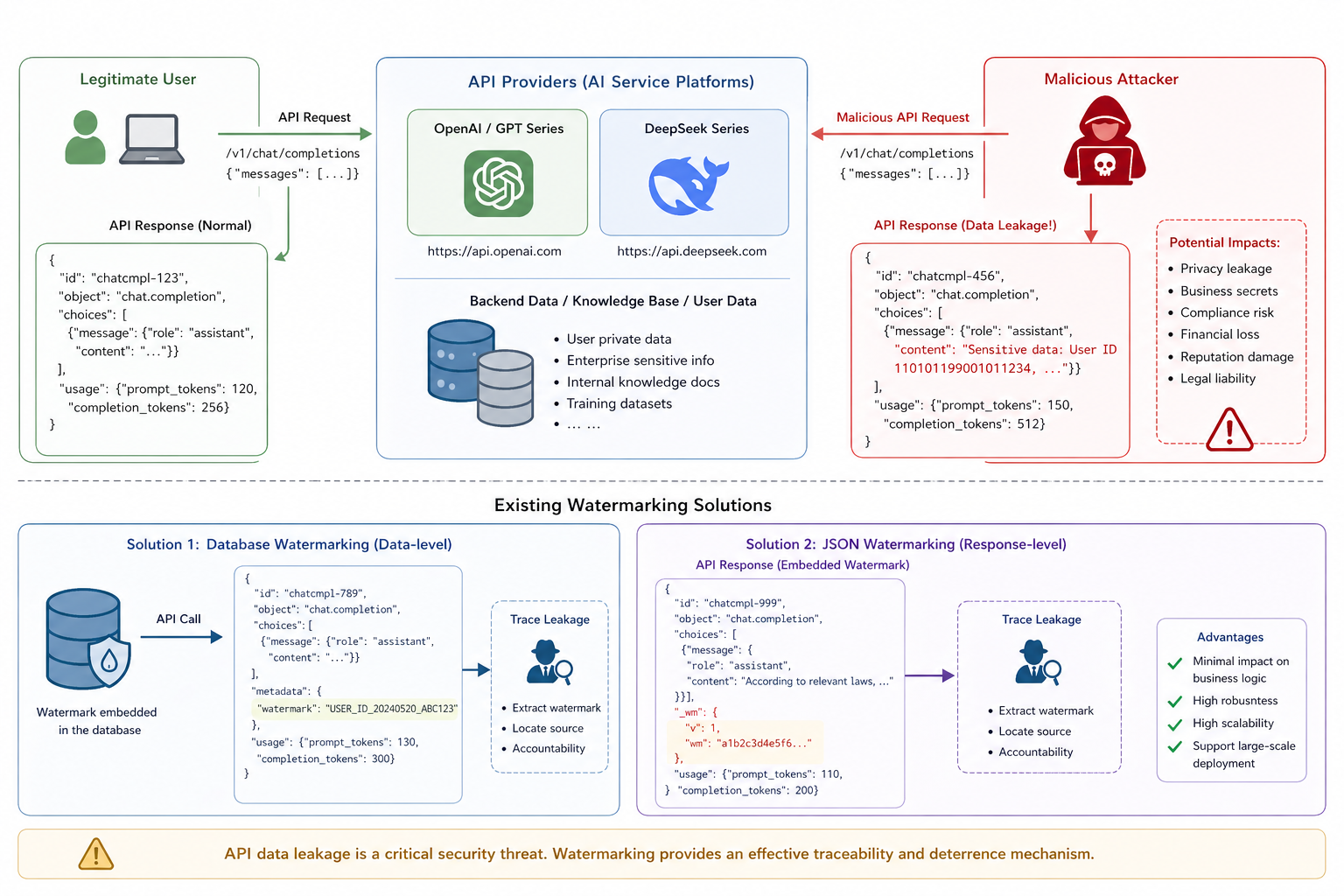}
	\caption{API data leakage risks and watermarking-based solutions.}
	\label{sec1}
\end{figure*}

A common solution is to embed watermark data into the database, and then use that watermarked data during API interactions \cite{halder2010watermarking, agrawal2002watermarking, shehab2008watermarking, hu2018new, ren2023robust, iftikhar2015rrw}. Another approach is to modify semi-structured JSON data in API responses to embed watermarks \cite{florescu1999performance, he2020semi, constantin2024watermark, zhao2024robust}. As shown in Fig.~\ref{sec1}, these approaches enable tracing the leakage source by embedding identifiable information either in backend data or API responses.

Sadly, adding watermarks to a database can interfere with normal business operations. It often forces developers to change the business code \cite{kumar2020recent, gort2021semantic, chen2022reversible, gort2025robust}. Modifying the semi-structured data \cite{florescu1999performance, batra2018comparative} returned by an API also brings uncertainty to the business. Even a small change, such as a shift in decimal places, may cause business errors \cite{bart2014transforming}. In short, both types of solutions face big challenges in practice because they require changes to existing business code \cite{gort2025robust}. Business systems are often unwilling to make such changes \cite{gartner_data_masking_2024}.

In recent years, some zero-watermarking methods have been proposed to avoid data distortion. For example, Zhang et al. \cite{ji2026phimark} proposed PhiMark, a zero-distortion robust database watermarking scheme using a partition-and-mask approach. Tang et al. \cite{tang2021pkmark} proposed PKMark, a zero-distortion blind reversible scheme based on primary key marking. Han et al. \cite{han2024distortion} proposed a distortion-free scheme using auxiliary data encoding. Zhao et al. \cite{zhao2024robust} used a robust solution distribution-based zero-watermarking method to trace semi-structured power data. Other zero-watermarking schemes \cite{ren2023zero, xu2025zero, bhattacharya2024distortion, jiao2025enhancing} also rely on feature libraries. However, the main drawback of these methods is that they need to maintain a trusted third-party feature library for watermark extraction.

We surveyed watermarking methods for API interfaces in recent years, as shown in Table~\ref{tab:survey}. There are three types of methods that can protect API responses: modifying structured data \cite{hu2018new, ren2023robust, li2022secure, li2024consistency, li2022database, iftikhar2015rrw, chen2022reversible, jawad2013genetic, alghamdi2022novel, rani2022comparative, cai2023low, qi2024research}, modifying semi-structured data \cite{he2020semi, constantin2024watermark, zhao2024robust}, and zero-watermark \cite{ji2026phimark, ren2023zero, xu2025zero, bhattacharya2024distortion}. Besides these, frequency-based watermarking \cite{isler2024freqywm} and format-independent frameworks \cite{rani2022format} have also been proposed. Most studies focus on structured data such as databases and CSV files. Few studies focus on protecting semi-structured data like JSON and XML. Although existing schemes \cite{he2020semi, ji2026phimark} can be applied to semi-structured data, they have several weaknesses: they need a third-party feature library, they require changes to existing business code, and they affect normal business operations. These issues hinder the adoption of such technologies.

\begin{table*}[!htbp]
\caption{Comparison of Different Watermarking Methods}
\centering
\label{tab:survey}
\begin{tabular}{cccc}
\hline
Method Classification & Requires Business Code Change & Requires Third-Party Library & Affects Business Usability \\ \hline
Modify structured data\cite{hu2018new, ren2023robust, li2022secure, li2024consistency, li2022database}      & $\checkmark$                        & $\times$                           & $\checkmark$                            \\
Modify semi-structured data\cite{he2020semi, constantin2024watermark} & $\checkmark$                        & $\times$                           & $\checkmark$                            \\
Zero-Watermark\cite{ji2026phimark, zhao2024robust, ren2023zero}            & $\times$                        & $\checkmark$                           & $\times$                            \\
Position-Encoding (Ours)         & $\times$                        & $\times$                           & $\times$                            \\ \hline
\end{tabular}
\end{table*}

Our proposed method based on position encoding avoids the above issues. The key originality of this work lies in observing a little-known redundancy — the order of JSON/XML elements returned by an API call. This is a subtle property that has been largely overlooked in the watermarking literature. This order redundancy gives us a novel insight: we can embed data into the order using position encoding, thereby achieving distortion-free watermarking that requires neither data value modification nor a third-party feature library. To the best of our knowledge, we are the first to exploit key-ordering redundancy for API response watermarking via a proxy gateway architecture. The main contributions of this paper are summarized as follows.

\begin{itemize}
	\item We are the first to identify and exploit key-ordering redundancy in semi-structured API responses as a watermarking channel, and propose a novel position encoding algorithm that embeds watermarks without changing any data values, thus leaving the original business completely unaffected.
	\item We design an original pluggable watermarking framework based on a proxy gateway for on-the-fly protection of API response data \cite{frankel2024dynamic}, requiring zero modification to existing server or client code.
	\item Compared with current mainstream schemes, our method is highly robust. It fully resists numerical tampering attacks and insertion attacks (100\% similarity), and can withstand certain levels of deletion attacks — all while maintaining negligible time overhead (below 0.65 ms).
\end{itemize}

The rest of this paper is organized as follows. Section~\ref{sec:watermarking_proxy_gateways} describes our proposed watermark proxy gateway, which proxies API responses from the server for watermark embedding. Section~\ref{sec:watermarking_scheme} presents the position encoding method used to embed watermarks. Section~\ref{exp} presents and discusses our experimental results. Section~\ref{lim} discusses the limitations of our scheme and future work. Finally, Section~\ref{con} concludes the paper.

\section{Watermarking Proxy Gateway}
\label{sec:watermarking_proxy_gateways}

\begin{figure}[!htbp]
\centering
\includegraphics[width=0.5\textwidth]{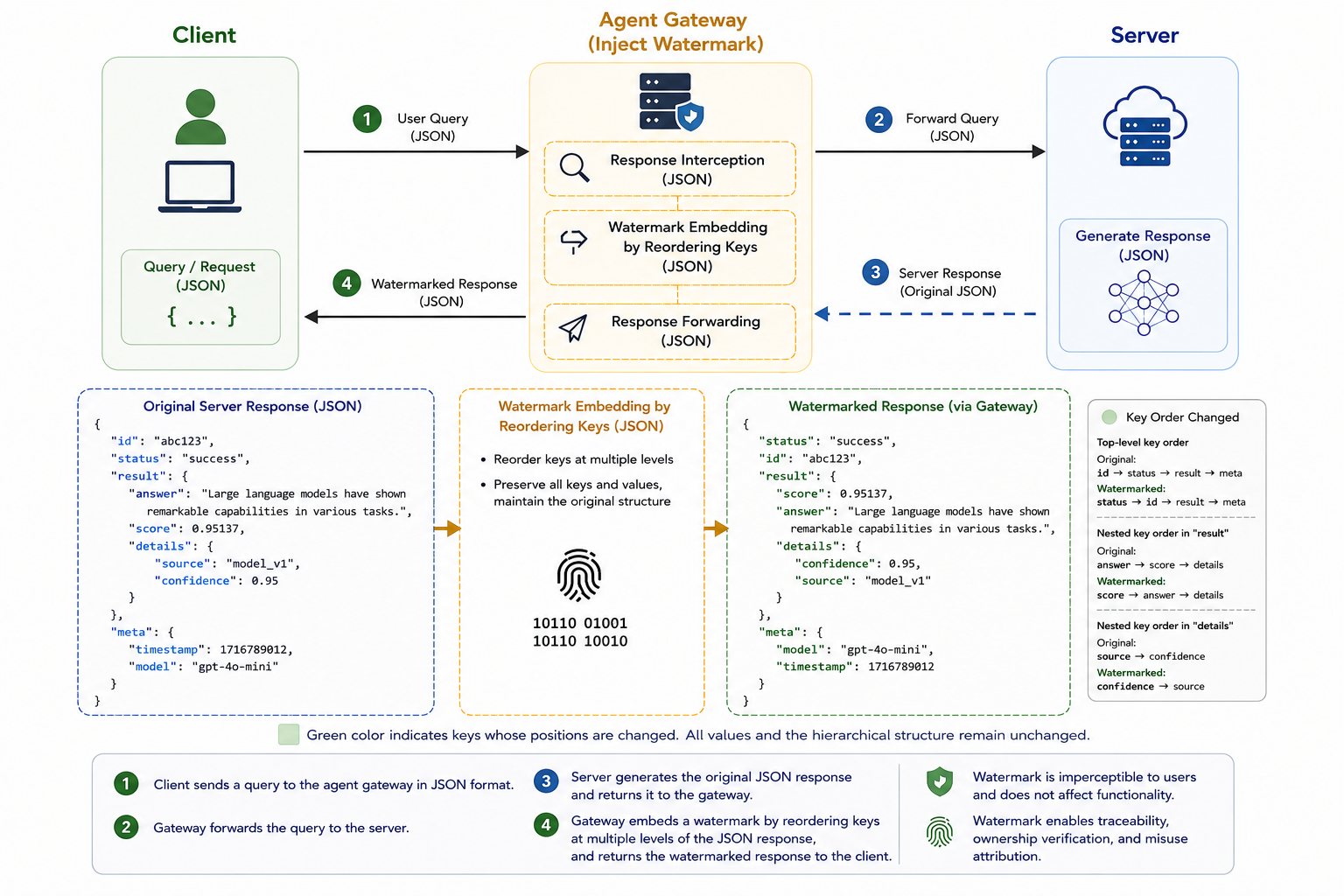}
\caption{Proxy gateway positioned between API server and client.}
\label{fig:proxy_gateway}
\end{figure}

We use a proxy gateway to embed watermarks into API responses. As illustrated in Fig.~\ref{fig:proxy_gateway}, the gateway is deployed between the API server and the client, intercepting each response before it reaches the client. This design is inspired by dynamic watermarking solutions for web content \cite{frankel2024dynamic} and distortion‑free watermarking for relational databases \cite{bhattacharya2024distortion}.

Upon receiving a response from the server, the gateway extracts the JSON data and applies the position encoding method described in Section~\ref{sec:watermarking_scheme} to embed the watermark. Interestingly, the gateway operates transparently: neither the server nor the client is modified. The server continues to function as usual, while the client receives the watermarked response without any awareness of the change. This design preserves the original API logic and requires no modifications to business code.

\section{Watermarking Scheme}
\label{sec:watermarking_scheme}

\begin{figure*}[!htbp]
	\centering
	\includegraphics[width=7 in]{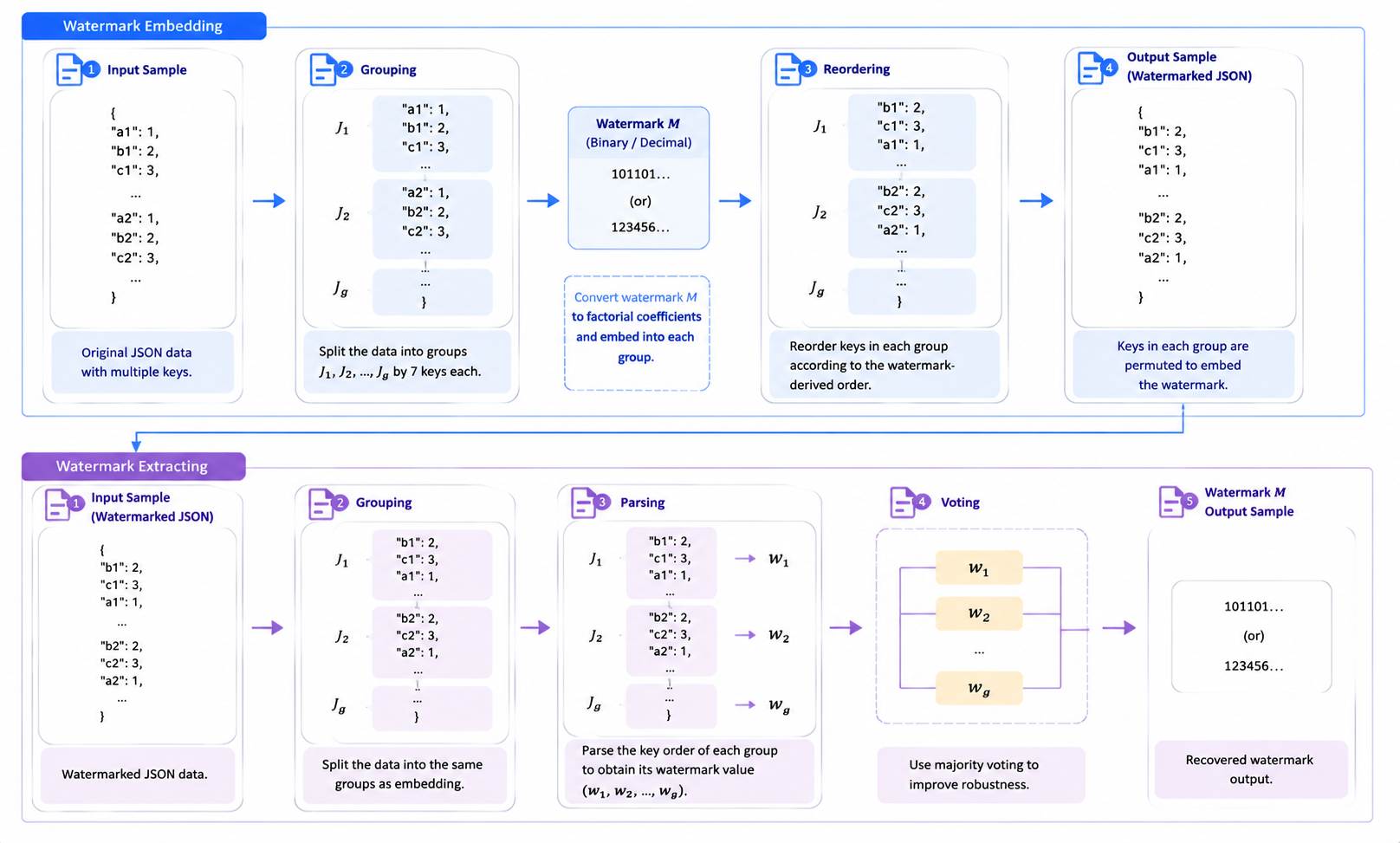}
	\caption{Overview of PEMark watermark embedding and extraction process.}
	\label{fig:pemark_overview}
\end{figure*}

This section introduces PEMark (Position Encoding-based Watermarking), a scheme that leverages the permutation characteristics of keys in semi-structured data to embed watermarks. We take JSON as an example to illustrate how position encoding works on semi-structured data. The entire scheme consists of two core components: embedding and extraction, as illustrated in Fig.~\ref{fig:pemark_overview}.

\subsection{Watermark Embedding}

The essence of watermark embedding is to convert a watermark value into a specific order of keys in JSON data. The embedding process is shown in Fig.~\ref{fig:pemark_overview}.

\subsubsection{Data Grouping}

For a JSON object $J$ that contains $N$ keys, denoted as $\{j_1, j_2, ..., j_N\}$, we group $J$ into $g$ groups: $J_1, J_2, ..., J_g$. This process can be expressed as:

\begin{equation}
J = \{J_1, J_2, ..., J_g\}
\label{eq:grouping}
\end{equation}
where $g$ is the number of groups, and each group $J_i$ contains \(g = \lfloor N/T \rfloor\) keys.

In each group $J_i$, the total number of possible permutations is $n!$. This means we have $n!$ different orders that can be used in $J_i$. Each group $J_i$ is used to embed one complete watermarks.

A capacity threshold $T$ is also needed for the watermark, because each group $J_i$ needs enough keys to carry a watermark of a certain length. The values of $g$ and $T$ balance watermark capacity against robustness. We discuss this trade-off in Section~\ref{sec:capacity_experiment}.

\subsubsection{Factorial Decomposition}

Given a binary watermark sequence $M={m_1,m_2,...,m_x}$, The value of $M$ is decomposed into a sequence of factorial coefficients through successive division with remainder as formula (\ref{eq:FD}). For the capacity threshold \(T\), we compute coefficients \(a_1, a_2, \ldots, a_{T-1}\) satisfying:

\begin{equation}
M = \sum_{j=1}^{T-1} a_j \times (T-j)!
\label{eq:FD}
\end{equation}
where each factorial coefficient \(a_j\) ranges in \([0, T-j-1]\) and corresponds to the index of the selected key in the remaining key set at each step.

\subsubsection{Group-based Reordering}

\begin{figure*}[!htbp]
	\centering
	\includegraphics[width=1\textwidth]{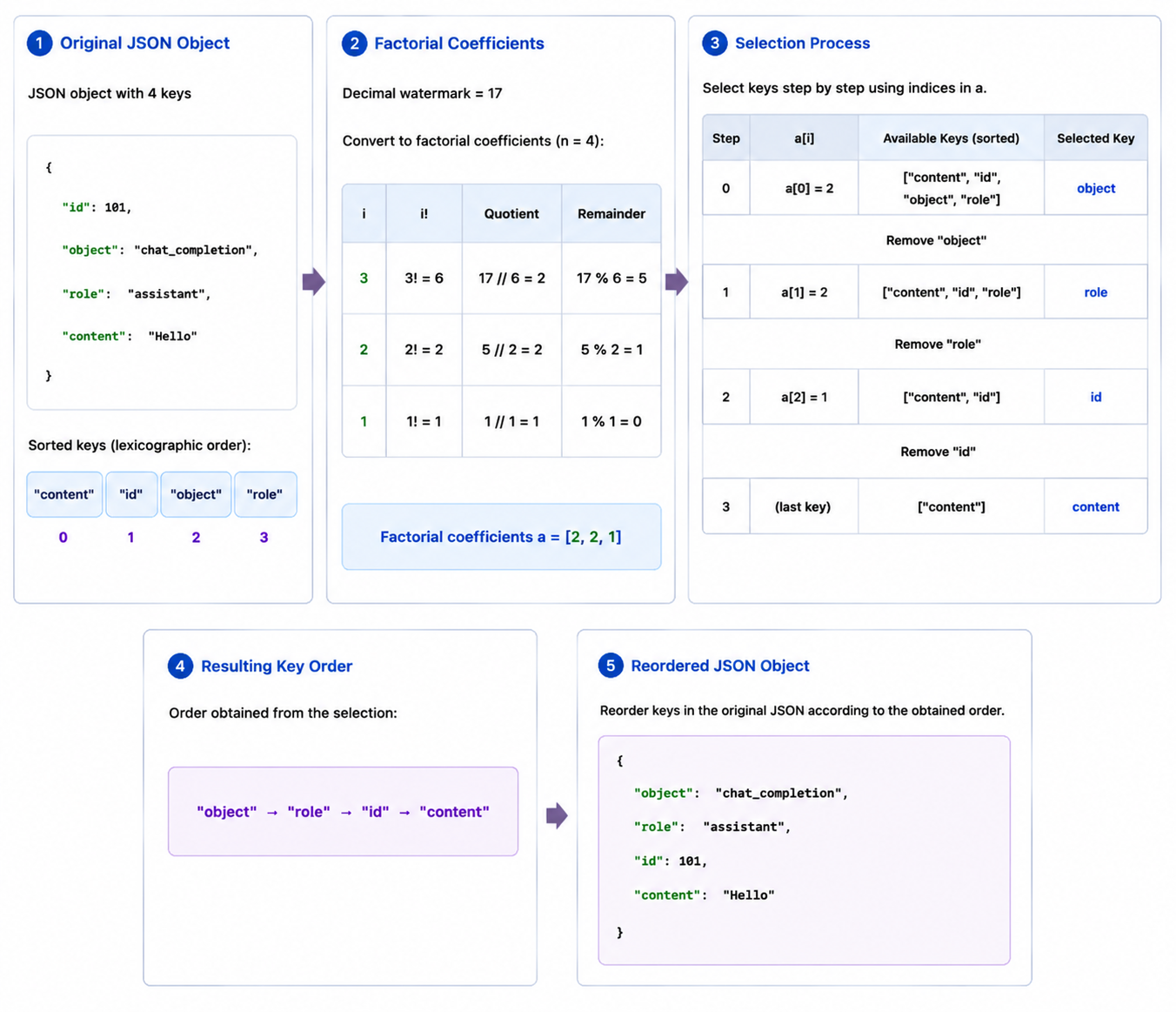}
	\caption{Key position determination inside a group during reordering.}
	\label{fig:group_reorder}
\end{figure*}

Fig.~\ref{fig:group_reorder} illustrates how the coefficient sequence from factorial decomposition determines the key order within a group.  The coefficients \((a_1, a_2, \ldots, a_{T-1})\) constitute a \emph{Lehmer code} \cite{lehmer1960teaching} --- a bijection between integers \([0, T!-1]\) and all \(T!\) permutations.  A Lehmer code \((L_1, \ldots, L_T)\) satisfies \(0 \leq L_k \leq T-k\), where \(L_k\) counts how many remaining elements are smaller than the current pick; our final coefficient \(a_T\) is implicitly zero.

For a group \(J_i\) with keys \(\{k_1, \ldots, k_T\}\), let the sorted key sequence be:

\begin{equation}
\mathbf{K}^{(0)} = \bigl(k_{(1)}, k_{(2)}, \ldots, k_{(T)}\bigr), \quad
k_{(1)} < k_{(2)} < \cdots < k_{(T)},
\label{eq:sorted_keys}
\end{equation}
where the ordering is lexicographic.  The watermarked permutation \(\boldsymbol{\pi} = (\pi_1, \pi_2, \ldots, \pi_T)\) is then constructed recursively:

\begin{equation}
\left\{ \begin{array}{l}
	\pi _j=\mathbf{K}^{\left( j-1 \right)}\left[ a_j \right]\\
	\mathbf{K}^{\left( j \right)}=\mathbf{K}^{\left( j-1 \right)}\setminus \{\pi _j\}\\
	\pi _T=\mathbf{K}^{\left( T-1 \right)}\left[ 0 \right]\\
\end{array} \right. \ \ \ ,\ j=1,2,...,T-1
\label{eq:embed_perm}
\end{equation}
where \(\mathbf{K}^{(j-1)}[a_j]\) denotes the element at index \(a_j\) (zero-based) in the ordered sequence \(\mathbf{K}^{(j-1)}\), and \(\setminus\) denotes set removal.  The constraint \(0 \leq a_j \leq T-j-1\) guarantees that the index is always valid.  The resulting watermarked group is:

\begin{equation}
J_w^{(i)} = \bigl\{ (\pi_1, v_{\pi_1}), (\pi_2, v_{\pi_2}), \ldots, (\pi_T, v_{\pi_T}) \bigr\}.
\label{eq:watermarked_group}
\end{equation}

After all \(g\) groups are processed, any remaining keys (those not forming a complete group) are appended in their original order.  These leftover keys carry no watermark information.

\subsection{Watermark Extraction}

Extraction recovers the watermark by inverting the permutation process.  Given the watermarked data \(J_w\), the group count is \(\lfloor N'/T \rfloor\) where \(N'\) is the total key count.  Only complete groups are used.

For a watermarked group with observed key order \((q_1, q_2, \ldots, q_T)\), the coefficients are reconstructed by reversing (\ref{eq:embed_perm}):

\begin{equation}
\left\{ \begin{array}{l}
	\overset{\left( 0 \right)}{\widehat{\mathbf{K}}}=\text{lex\_sort}\bigl\{ \{q_1,...,q_T\} \bigr)\\
	\widehat{a}_j=\text{index\_of}\bigl( q_j,\;\overset{\left( j-1 \right)}{\widehat{\mathbf{K}}} \bigr)\\
	\overset{\left( j \right)}{\widehat{\mathbf{K}}}=\overset{\left( j-1 \right)}{\widehat{\mathbf{K}}}\setminus \{q_j\}\\
\end{array} \right. \ \ \ ,\ j=1,2,...,T-1
\label{eq:extract_coeff}
\end{equation}

Each \(\widehat{a}_j\) is the zero-based position of key \(q_j\) in the current sorted remainder, which exactly matches the original coefficient \(a_j\) when no keys have been deleted or reordered by an attacker.  The watermark integer is then recovered from the factorial series:

\begin{equation}
\widehat{M} = \sum_{j=1}^{T-1} \widehat{a}_j \times (T-j)!,
\label{eq:extract_sum}
\end{equation}

and is converted to a binary sequence of length \(L\):

\begin{equation}
\widehat{W} = \text{bin}\bigl(\widehat{M},\; L\bigr),
\label{eq:bin_convert}
\end{equation}

where \(\text{bin}(x, L)\) returns the \(L\)-bit binary representation of \(x\), padding with leading zeros as needed.

Because embedding and extraction share the same lexicographic baseline and factorial mapping, the recovered coefficients satisfy \(\widehat{a}_j = a_j\) in the absence of attacks, guaranteeing zero-bit-error extraction.

Finally, majority voting is applied across all \(G\) groups.  For each bit position \(j\):

\begin{equation}
w_{\text{final}}[j] = \operatorname{mode}\bigl(w_1[j], w_2[j], \ldots, w_G[j]\bigr),
\label{eq:voting}
\end{equation}

where \(w_g[j]\) is the \(j\)-th bit from group \(g\).  This mechanism ensures that the correct watermark is recoverable as long as a majority of groups remain intact.

\section{Experiment}
\label{exp}

This section introduces evaluation metrics (\ref{exp}-A), experimental setup (\ref{exp}-B), capacity experiment (\ref{exp}-C), time overhead analysis (\ref{exp}-D), robustness comparison with baseline methods (\ref{exp}-E), and testing in real-world systems (\ref{exp}-F).

\subsection{Experimental Evaluation}

Three metrics are adopted to evaluate the performance of the proposed method: time overhead, attack intensity, and watermark similarity.

\subsubsection{Time Overhead}
Time overhead refers to the execution time required to embed the watermark into semi-structured data and extract it from the watermarked data. Specifically, it includes:

\begin{itemize}
    \item \textbf{Embedding Time}: The total time required to generate the watermarked version from the original semi-structured data.
    \item \textbf{Extraction Time}: The total time required to recover the watermark information from the watermarked semi-structured data.
\end{itemize}

Time overhead is measured in milliseconds (ms). A larger time overhead indicates a greater negative impact on normal business processes for the given dataset, i.e., lower operational efficiency of the method.

\subsubsection{Attack Intensity}

Attack intensity quantifies the degree of malicious modification imposed on semi-structured data. Given a semi-structured data sample containing \(N\) key-value pairs, the attacker can perform one of the following three operations:

\begin{itemize}
    \item \textbf{Deletion}: Randomly remove \(M\) key-value pairs.
    \item \textbf{Tampering}: Randomly modify the values of \(M\) key-value pairs.
    \item \textbf{Insertion}: Randomly add \(M\) new key-value pairs.
\end{itemize}

Attack intensity is defined as follows:

\begin{equation}
\text{Attack Intensity} = \frac{M}{N} \times 100\%
\end{equation}

where \(M\) denotes the number of key-value pairs that are deleted, tampered, or inserted, and \(N\) denotes the total number of key-value pairs in the original semi-structured data. As defined in (1), a higher attack intensity indicates more severe damage to the semi-structured data, making correct watermark extraction more challenging. In this experiment, we set the attack intensity range from 0\% to 50\% to evaluate the robustness of the method under different levels of destruction.

\subsubsection{Watermark Similarity}
The watermark similarity is calculated as:

\begin{equation}
\text{Watermark Similarity} = \frac{1}{L} \sum_{i=1}^{L} I(w_i = w'_i) \times 100\%
\end{equation}

where \(I(\cdot)\) is the indicator function that takes the value 1 when the condition holds and 0 otherwise. According to (2), a higher watermark similarity indicates less damage to the embedded watermark, implying a greater probability of successfully recovering the correct original watermark under error correction coding.

\subsection{Experimental Setup}

\subsubsection{Testbed}
The watermarking system was built upon the OpenResty\footnote{\url{https://github.com/openresty/openresty}}  Proxy framework for HTTP request/response interception. Each experiment was repeated ten times with random seeds, and the average results are reported.

\subsubsection{Baseline Methods}
We compared our method with three state-of-the-art semi-structured data watermarking schemes: He et al.\cite{hu2018new}, which embeds watermarks by modifying the values of key-value pairs; Phimark \cite{ji2026phimark}, a zero-distortion watermarking scheme for relational data; and Zhao et al. \cite{zhao2024robust}, a zero-watermarking scheme based on soliton distribution for semi-structured power data.

\subsection{Datasets}
Due to data sensitivity, we used desensitized data from the power industry to construct the API server-side datasets for our experiments. Table~\ref{tab:dataset_config} summarizes the configuration of the nine datasets used in our evaluation.

\begin{itemize}
    \item \textbf{Number of Keys}: The count of key-value pairs in each JSON object.
    \item \textbf{Max Nesting Depth}: The deepest nested level within the JSON structure. Depth 1 means no nesting (flat key-value pairs).
    \item \textbf{Nesting Probability}: The chance that a given key contains a nested object instead of a primitive value.
\end{itemize}

These settings cover a wide range of real-world semi-structured data and help us test PEMark under different structural conditions.

\begin{table}[!htbp]
\centering
\caption{Configuration of constructed datasets}
\label{tab:dataset_config}
\begin{tabular}{cccc}
\hline
ID & Number of Keys & Max Nesting Depth & Nesting Probability \\
\hline
1 & 5--25 & 1 & 0 \\
2 & 5--25 & 3 & 0.3 \\
3 & 5--25 & 3 & 0.7 \\
4 & 50--250 & 1 & 0 \\
5 & 50--250 & 3 & 0.3 \\
6 & 50--250 & 3 & 0.7 \\
7 & 50 & 3 & 0.3 \\
8 & 100 & 3 & 0.3 \\
9 & 200 & 3 & 0.3 \\
\hline
\end{tabular}
\end{table}

\subsection{Capacity experiment}
\label{sec:capacity_experiment}

The capacity of PEMark is determined by the number of keys in each group and the factorial decomposition method. This section analyzes the relationship between watermark length, capacity threshold \(T\), and the number of groups \(g\).

\begin{figure}[!htbp]
\centering
\includegraphics[width=0.45\textwidth]{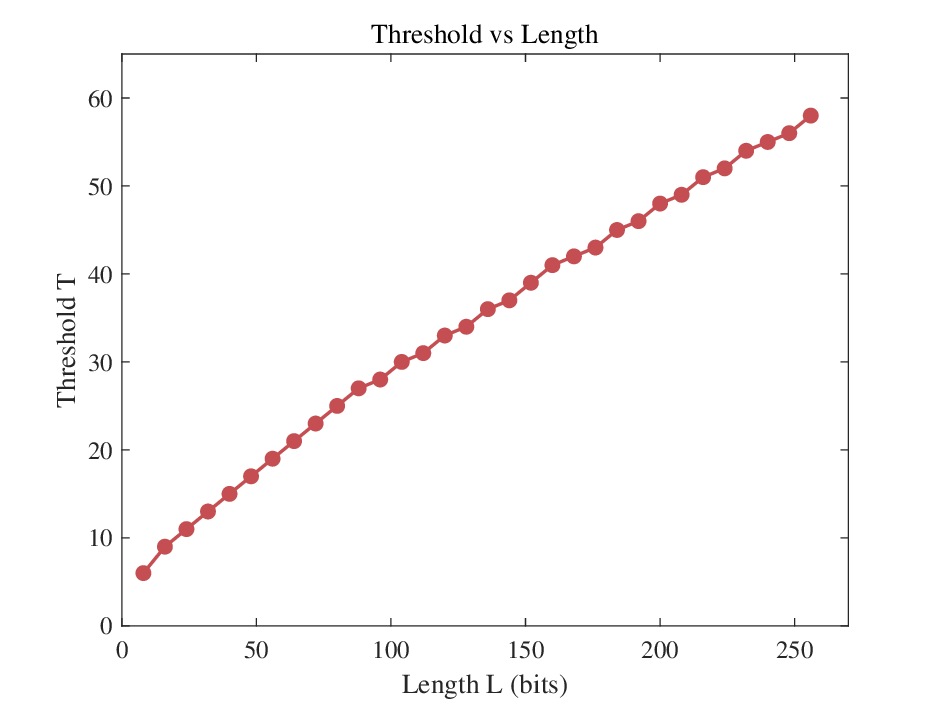}
\caption{Watermark length \(L\) vs. capacity threshold \(T\)}
\label{fig:capacity}
\end{figure}

\subsubsection{Watermark Length vs. Capacity Threshold}

A longer watermark requires more permutations to encode all possible values. The capacity threshold \(T\) is the minimum number of keys needed in a group to embed a watermark of length \(L\) bits. The condition is given by:

\begin{equation}
2^{L} \leq T!
\label{eq:capacity_condition}
\end{equation}

Fig.~\ref{fig:capacity} shows the relationship between \(L\) and \(T\). For instance, a 64-bit watermark requires \(T = 21\) keys, while 128-bit watermarks require \(T = 34\) keys.

\subsubsection{Robustness vs. Number of Groups}

The number of groups $g$ affects robustness. A larger \(g\) means the watermark is repeated in more groups. This gives more votes during extraction. As a result, the method can tolerate more group damage.

\subsubsection{Handling Insufficient Keys}

When a group has fewer than \(T\) keys, we cannot directly embed the full watermark. Our solution is to add fake key-value pairs to that group.

\subsection{Time Overhead Analysis}

\begin{figure*}[!htbp]
    \centering
    \includegraphics[width=0.3\textwidth]{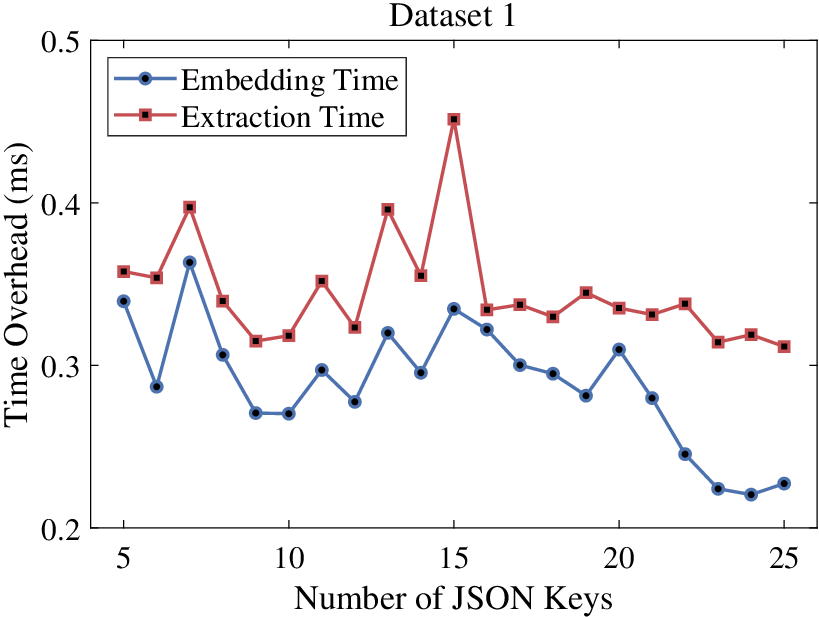}
    \hfill
    \includegraphics[width=0.3\textwidth]{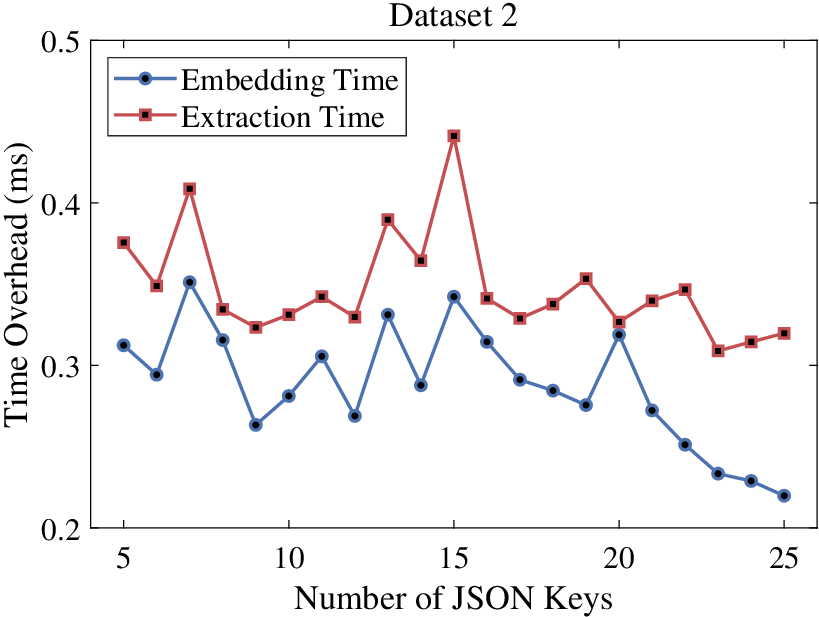}
    \hfill
    \includegraphics[width=0.3\textwidth]{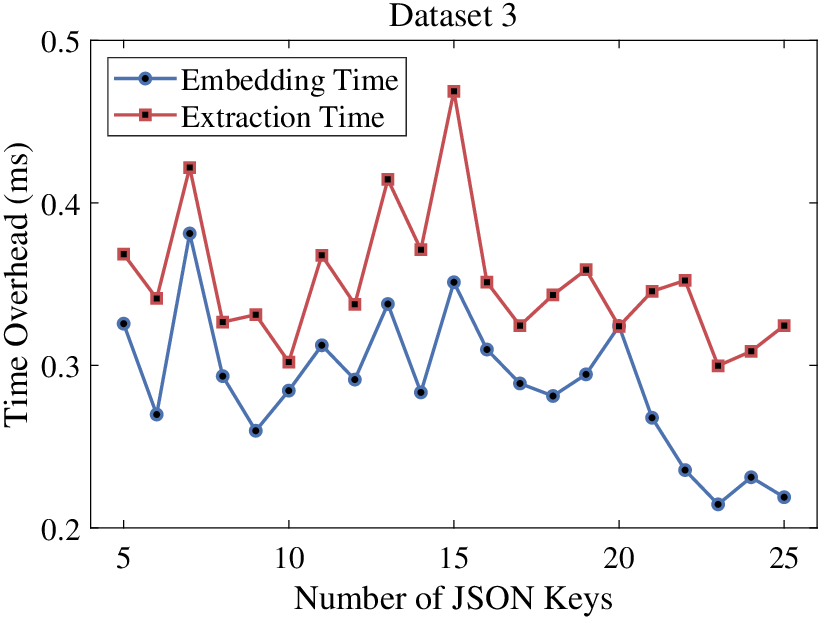}
    
    \vspace{0.5cm}
    
    \includegraphics[width=0.3\textwidth]{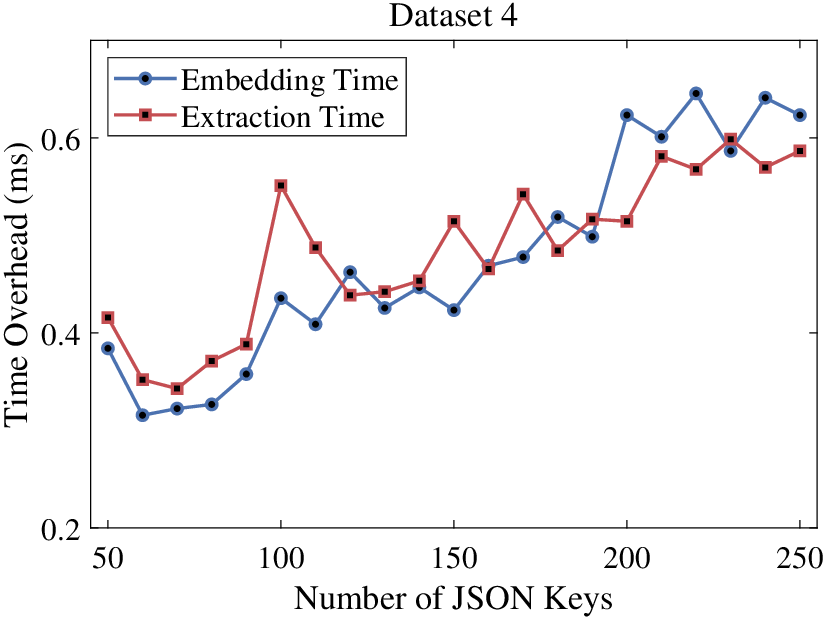}
    \hfill
    \includegraphics[width=0.3\textwidth]{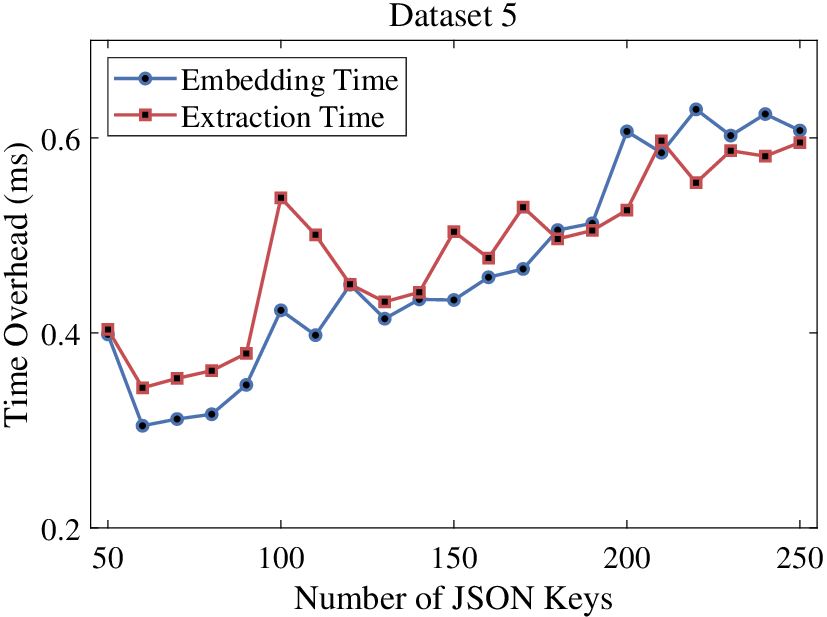}
    \hfill
    \includegraphics[width=0.3\textwidth]{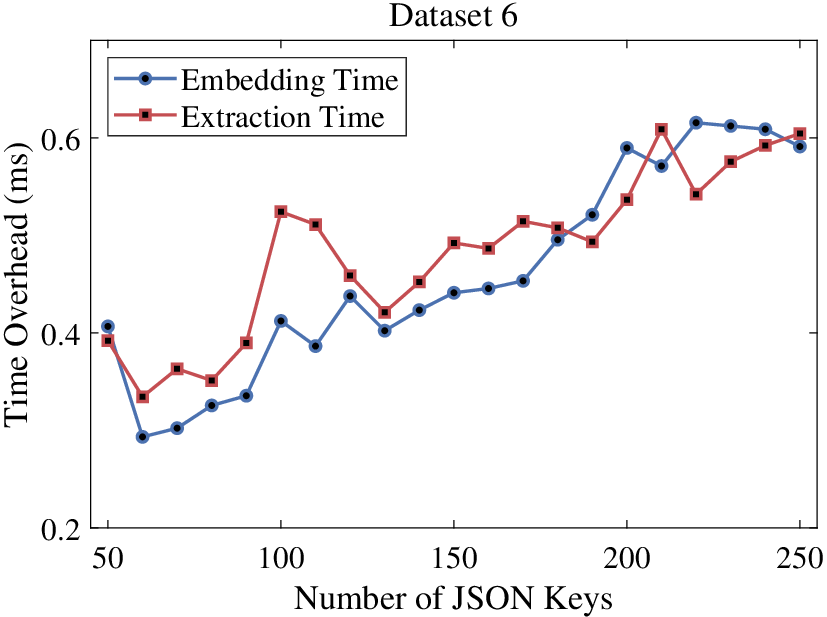}
    
    \caption{Time overhead comparison of embedding and extraction on six datasets.}
    \label{fig:time_results}
\end{figure*}

We measured embedding and extraction time on six datasets with different sizes. Fig.~\ref{fig:time_results} shows the results.

As the number of keys increases, both embedding and extraction time increase. For datasets with 5--25 keys, the time ranges from 0.22~ms to 0.47~ms. For datasets with 50--250 keys, the time ranges from 0.27~ms to 0.65~ms.

PEMark keeps time overhead low. Even on datasets with 250 keys and high nesting complexity, the time stays between 0.3~ms and 0.6~ms.

\subsection{Robustness Comparison with Baseline Methods}

\begin{figure*}[!htbp]
    \centering
    \includegraphics[width=0.3\textwidth]{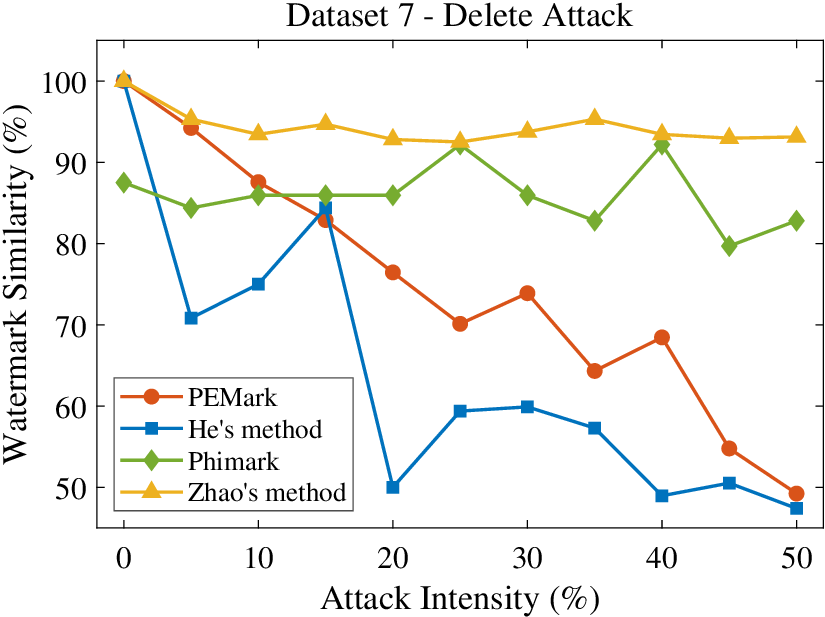}
    \hfill
    \includegraphics[width=0.3\textwidth]{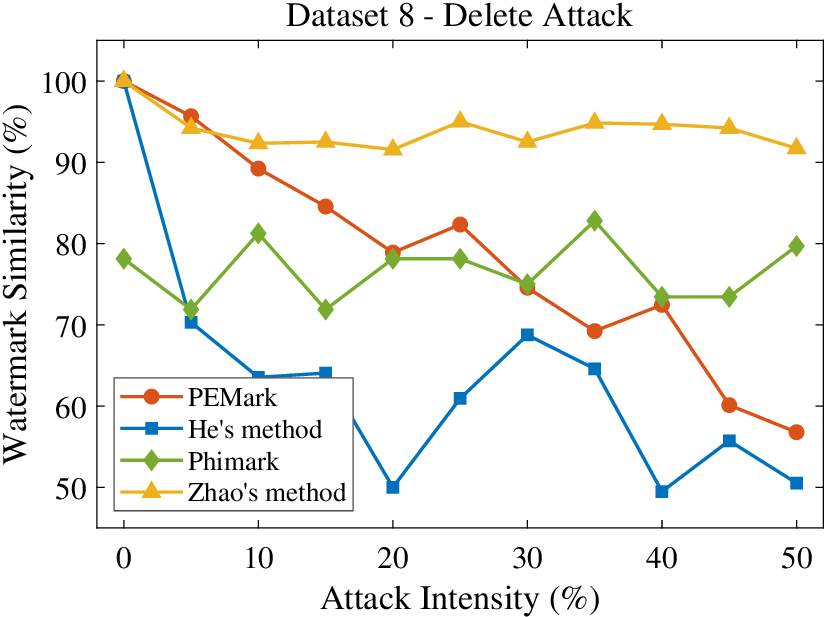}
    \hfill
    \includegraphics[width=0.3\textwidth]{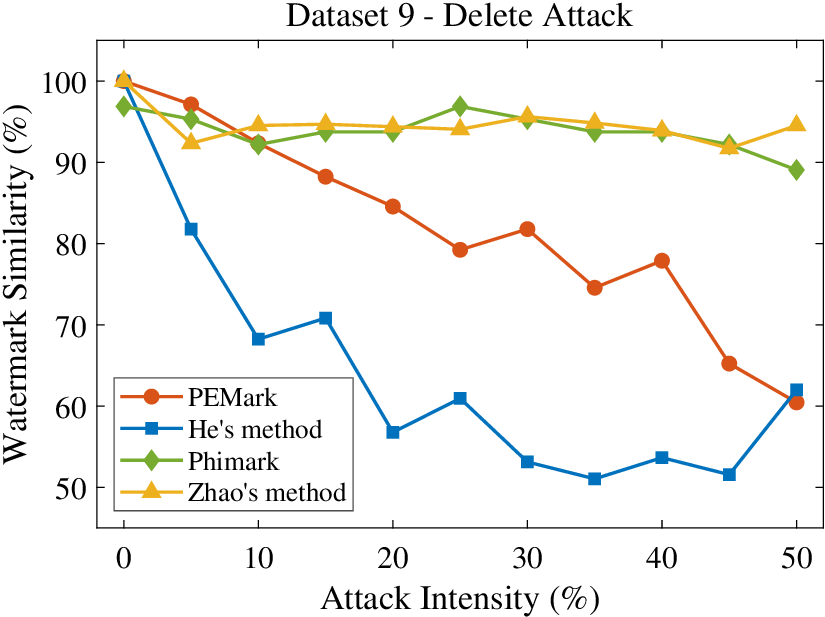}
    
    \vspace{0.5cm}
    
    \includegraphics[width=0.3\textwidth]{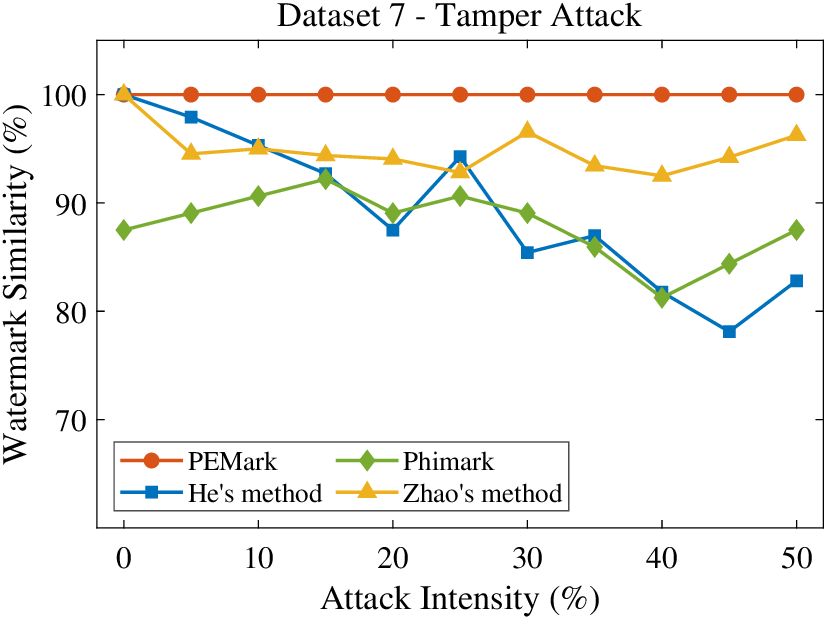}
    \hfill
    \includegraphics[width=0.3\textwidth]{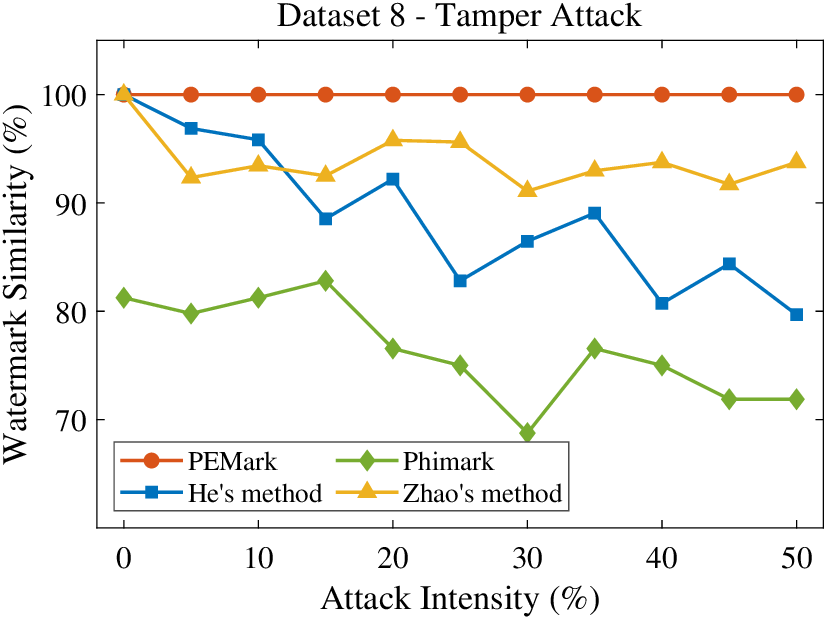}
    \hfill
    \includegraphics[width=0.3\textwidth]{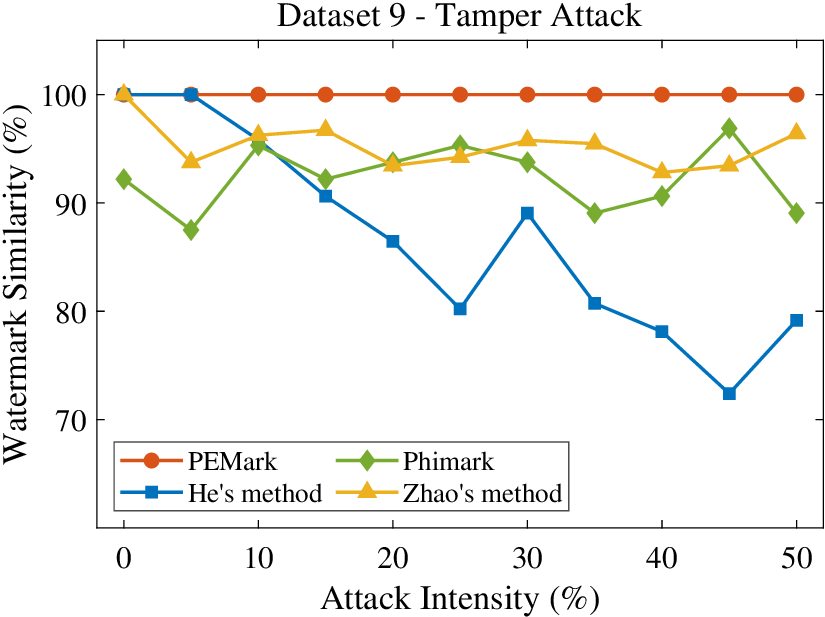}
    
    \vspace{0.5cm}
    
    \includegraphics[width=0.3\textwidth]{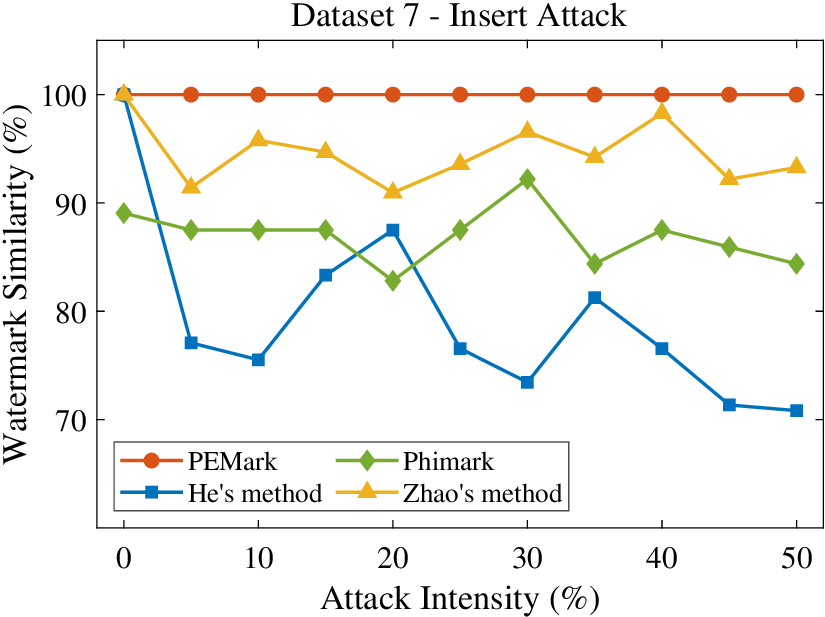}
    \hfill
    \includegraphics[width=0.3\textwidth]{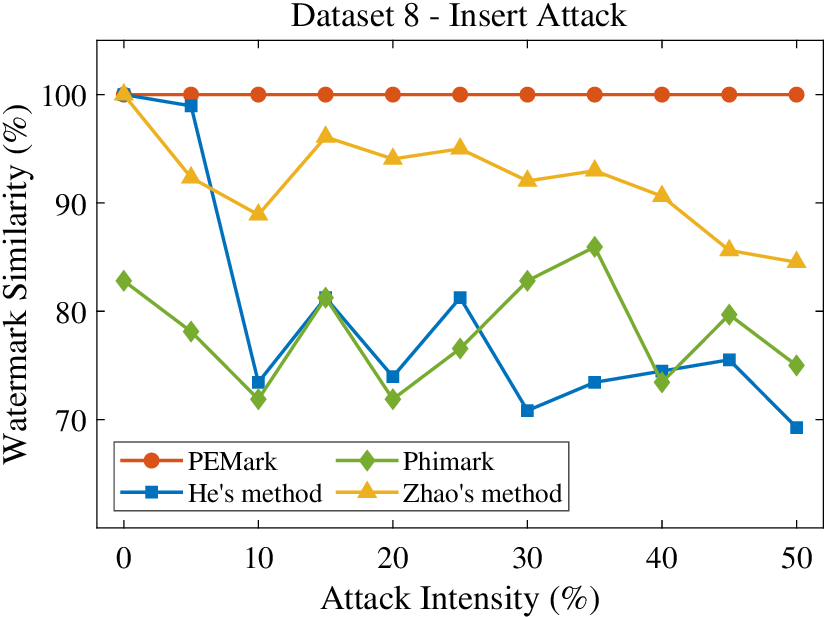}
    \hfill
    \includegraphics[width=0.3\textwidth]{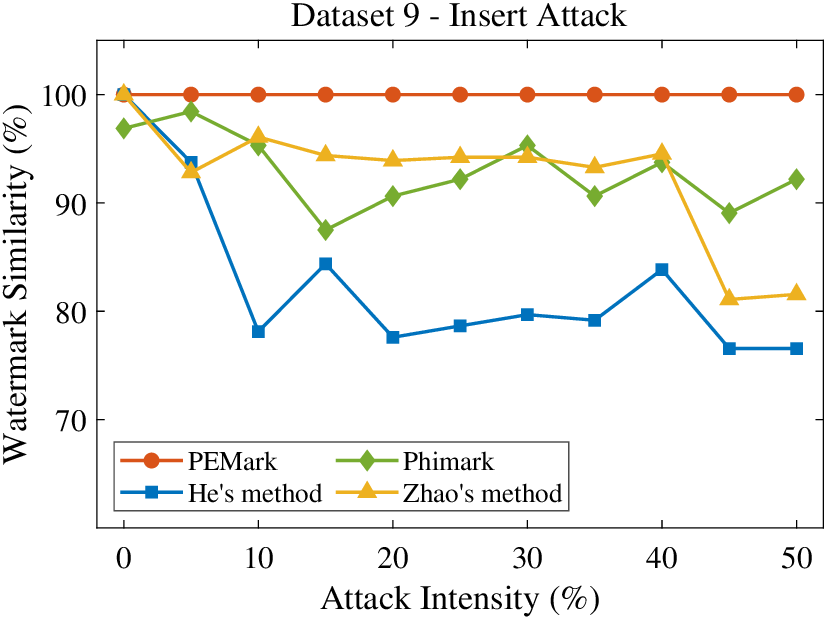}
    
    \caption{Robustness comparison of PEMark with baseline methods under different attack types.}
    \label{fig:robustness_results}
\end{figure*}

We compared PEMark with three methods: He's method\cite{hu2018new}, Phimark\cite{ji2026phimark}, and Zhao's method\cite{zhao2024robust}. We tested three attacks: deletion, tampering, and insertion. We used three datasets: Dataset 7 (50 keys), Dataset 8 (100 keys), and Dataset 9 (200 keys). Fig.~\ref{fig:robustness_results} shows the results.

\subsubsection{Delete Attack}

PEMark keeps similarity above 94\% when attack intensity is below 15\%. At 50\% attack intensity, PEMark gives 49.23\%-60.45\% similarity. He's method drops to 47.40\%-61.98\% with large fluctuations. Phimark stays at 75.10\%-96.75\%. Zhao's method keeps above 91\%.

Delete attacks break the order of keys. So PEMark does not beat zero-watermark methods like Zhao's method. But PEMark works better than He's method with more stable results.

\subsubsection{Tamper Attack}

PEMark keeps 100\% similarity under all attacks on all datasets. He's method drops to 79.17\%-82.81\% at 50\% attack. Phimark stays at 71.32\%-96.88\%. Zhao's method keeps above 91\%.

PEMark is perfect against tamper attacks. Our method does not change key values. It uses position encoding. So value changes do not affect the watermark. No other method achieves this.

\subsubsection{Insert Attack}

PEMark keeps 100\% similarity under all attacks on all datasets. He's method drops to 69.27\%-76.56\% at 50\% attack. Phimark stays at 72.42\%-98.04\%. Zhao's method drops to 81.56\%-84.53\%.

PEMark is also perfect against insert attacks. Two reasons explain this. First, insert attacks do not change existing keys that hold the watermark. Second, our voting mechanism rejects fake watermarks from new keys. Together, they make PEMark fully robust to insert attacks.

\subsection{Testing in Real-World System}
\label{sec:business_impact}

We tested PEMark on three real public APIs: DeepSeek\footnote{\url{https://api.deepseek.com/models}}, OpenAI\footnote{\url{https://api.openai.com/v1/models}}, and GitHub\footnote{\url{https://api.github.com/users/google}}. For each service, we measured the response time in two scenarios: (1) calling the API normally, and (2) applying PEMark to reorder the keys in the JSON response before forwarding it to the client.

\begin{figure}[!htbp]
\centering
\includegraphics[width=0.45\textwidth]{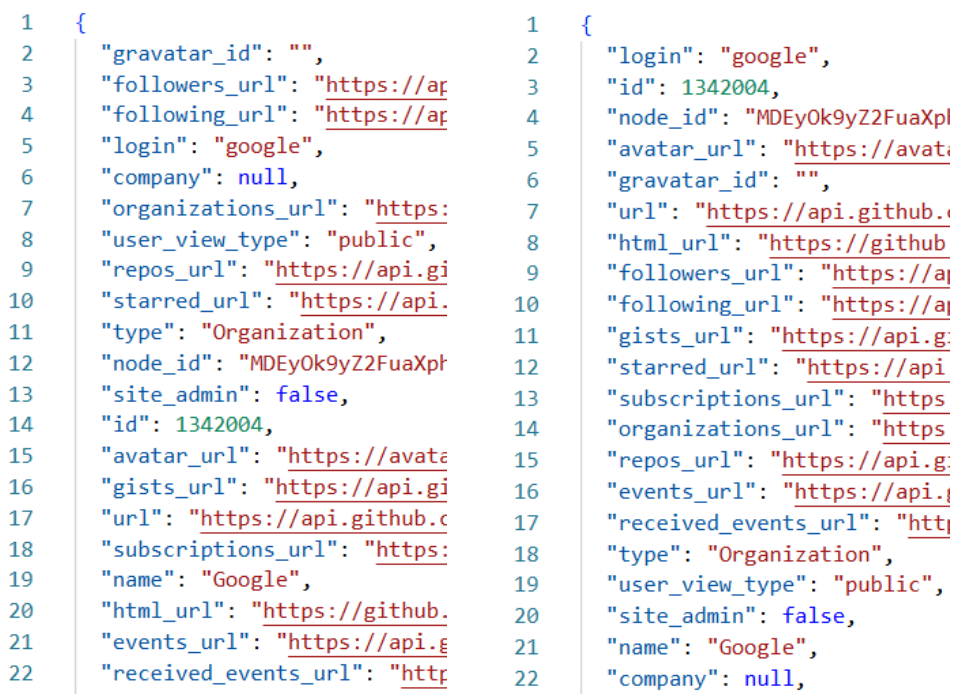}
\caption{Comparison of GitHub API JSON response before (left) and after PEMark processing (right).}
\label{fig:pemark_example}
\end{figure}

Fig.~\ref{fig:pemark_example} illustrates the effect of PEMark on the JSON response from the GitHub API. As shown, the semantic content remains unchanged, while the order of keys is rearranged according to the embedded watermark.

Table~\ref{tab:api_overhead} shows the results. The overhead of PEMark is consistent and low. All APIs remain fully usable after processing.

\begin{table}[!htbp]
\centering
\caption{API response time before and after PEMark processing.}
\label{tab:api_overhead}
\begin{tabular}{cccc}
\hline
API Service & Original (ms) & With PEMark (ms) & Overhead (ms) \\
\hline
DeepSeek & 543.24 & 563.19 & 19.95 \\
OpenAI & 933.92 & 952.47 & 18.55 \\
GitHub & 702.74 & 723.32 & 20.58 \\
\hline
\end{tabular}
\end{table}

PEMark does not change the API server or the logical data. Clients get the same data and work as before. The added latency is about 20 ms per request, which is acceptable for most real-world applications.
\section{Limitation}
\label{lim}

Our method has several limitations.

First, it does not work well under strong delete attacks. At 50\% attack strength, the similarity drops to 49\%-60\%. Delete attacks change the order of keys. This order is the base of our method. The topology of the watermark source significantly affects robustness \cite{gort2024study}. Future work can add error correction codes to fix this.

Second, our method uses a fixed key order. The keys are sorted from a to z. If an attacker knows this, they can change specific keys to break the watermark. In untrusted environments, watermark resilience to brute force attacks must be carefully considered \cite{gort2023relational}. One way to fix this is to use a secret key. The secret key randomizes the order. Only the sender and receiver know it.

Third, our method does not use deep nesting. It only looks at the top level of the JSON data. Nested objects have more space for watermarks. Future work can use this space to store more data or improve robustness.

We will address these limitations in future work.

\section{Conclusions}
\label{con}
We introduced PEMark, a novel position encoding-based watermarking scheme for API responses. The originality of this work is threefold: (1) we are the first to exploit key-ordering redundancy as a watermarking channel in semi-structured API responses; (2) we design a pluggable proxy-gateway architecture that requires zero modification to business code; and (3) we achieve distortion-free watermarking without any third-party feature library. Compared to existing methods, our scheme needs no third-party library and no business code changes. PEMark achieves perfect robustness against tamper and insert attacks (100.00\% similarity under all attacks). It also keeps 71\%-83\% similarity under 30\% delete attacks. Time overhead stays below 0.65 ms even for datasets with 250 keys. Future work will explore adaptive encoding for strong delete attacks and key-dependent ordering to stop targeted attacks.

\section*{Acknowledgment}

This work is supported by the Research on Key Technologies for Marking-Driven Detection and Traceability of Sensitive Electric Power Data Leakage. (No.52150025001H-146-ZN).

\bibliographystyle{IEEEtran}
\bibliography{bibRef}

\end{document}